\documentclass[twocolumn,showpacs]{revtex4}
\def\address{\affiliation}
\usepackage[dvips]{graphicx}
\begin{document}

\title{Thermal conductivity of the A$_x$BO$_2$ type layered oxides 
Na$_{0.77}$MnO$_2$ and LiCoO$_2$}

\author{Kazumi {\sc Takahata}$^{1}$ 
and Ichiro~{\sc Terasaki}$^{1,2}$ \footnote{E-mail address:
terra@mn.waseda.ac.jp}}

\address{
$^{1}$Department of Applied Physics, Waseda University,
3--4--1 Ohkubo, Shinjuku-ku, Tokyo, 169-8555, Japan\\
$^{2}$Precursory Research for Embryonic Science and
Technology, Japan Science Technology, Tokyo 108-0075, Japan
}

\date{\today}

\begin{abstract}
We prepared polycrystalline samples of the A$_x$BO$_2$ type  
layered oxides Na$_{0.77}$MnO$_2$ and LiCoO$_2$,
and measured thermal conductivity ($\kappa$) from 15 to 280 K. 
Na$_{0.77}$MnO$_2$ shows a low $\kappa$ of 15 mW/cmK at 280 K, 
while LiCoO$_2$ shows a much higher $\kappa$ of 40 mW/cmK at 280 K.
By comparing them with $\kappa$ of NaCo$_2$O$_4$,
we have found  that the low $\kappa$ of A$_x$BO$_2$
comes essentially from the A-ion deficiency,
not from the layered structure itself.
\end{abstract}



\maketitle


Recently, thermoelectric energy conversion \cite{Mahan} has attracted 
renewed interest from the viewpoint of environmental concerns.
In this respect, oxides are quite  desirable, because they are
stable in air and mostly harmless.
Owing to the low efficiency,
they have been thought to be unsuitable 
for thermoelectric applications, but
the layered cobalt oxide NaCo$_2$O$_4$ has been found to show
exceptionally high thermoelectric performance \cite{terasaki}.
Previously we found that NaCo$_2$O$_4$
has a low thermal conductivity ($\kappa$) of 15~mW/cmK at 280 K,
which was attributed to the short phonon mean free path 
arising from the Na deficiency \cite{takahata}.

An open question is whether the layered structure 
is effective for the $\kappa$ reduction in NaCo$_2$O$_4$ or not,
as in the case of the thermoelectric superlattice \cite{hicks,Chen,Simkin}.
Although the phonon-scattering theory \cite{Callaway1,Callaway2}
successfully fitted the data in the previous paper \cite{takahata}, 
we did not rule out the possibility of the layered-structure effects. 
The best way is perhaps to measure anisotropic $\kappa$ using
single crystals, but the crystals are too small and thin 
to measure $\kappa$ accurately \cite{comment}.
At least we can say that there is no reason to regard
NaCo$_2$O$_4$ as a superlattice on the electric properties.
The resistivity is anisotropic, but quantum confinement is 
unlikely to occur in the c-axis conduction \cite{terasaki}.

In order to distinguish the layered-structure effects from the
Na-deficiency effects,
we focus on Na$_{0.77}$MnO$_2$ and LiCoO$_2$, 
which have the same layered structure of A$_x$BO$_2$,
but have different A-ion deficiency (see Fig.1):
Na ions in NaCo$_2$O$_4$ (Na$_{0.5}$CoO$_2$) 
and Na$_{0.77}$MnO$_2$ are 50 \% and 23\% deficient, 
while Li ions in LiCoO$_2$ are nominally 100\% occupied.
Here we report on measurement and analysis 
of $\kappa$ of Na$_{0.77}$MnO$_2$ and LiCoO$_2$.

Polycrystalline samples of Na$_{0.77}$MnO$_2$ and LiCoO$_2$ were 
prepared through a solid-state reaction.
For Na$_{0.77}$MnO$_2$, starting powders of NaCO$_3$ and 
Mn$_3$O$_4$ were mixed and calcined at 500$^{\circ}$C for 12h.
The product was pressed into a pellet, and sintered at 
550$^{\circ}$C for 12h \cite{parant}.
For LiCoO$_2$, starting powders of LiCO$_3$
and Co$_3$O$_4$ were mixed and calcined at 450$^{\circ}$C for 12h.
The product was pressed into a pellet, and sintered 
at 900$^{\circ}$C for 24h \cite{delmas}.
X-ray diffraction revealed no impurity phases in the obtained
polycrystals, and resistivity measurement revealed that they
were highly insulating.
The thermal conductivity was measured using a steady-state technique
described previously \cite{takahata}.

\begin{figure}[h]
 \includegraphics[width=8cm,clip]{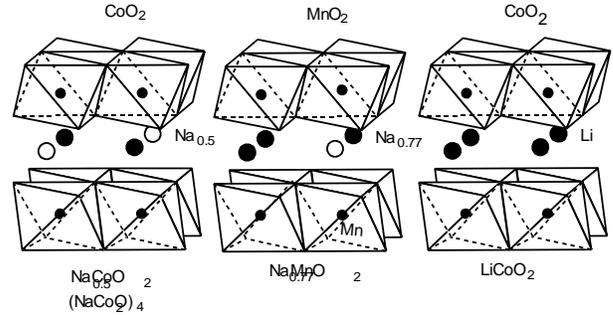}
 \caption{
 Crystal structures of Na$_{0.5}$CoO$_2$ (NaCo$_2$O$_4$),
 Na$_{0.77}$MnO$_2$ and LiCoO$_2$.
 The close and open circles represent the occupied and vacant ions,
 respectively. 
 }
\end{figure}

Figure 2 shows $\kappa$ of Na$_{0.77}$MnO$_2$ and LiCoO$_2$, 
together with $\kappa$ of NaCo$_2$O$_4$ from Ref.\cite{takahata}.
$\kappa$ of the Na-deficient oxide Na$_{0.77}$MnO$_2$ 
is as low as 15 mW/cmK at 280 K, 
and is slightly lower than $\kappa$ of NaCo$_2$O$_4$.
Since $\kappa$ of NaCo$_2$O$_4$ includes 
the electron thermal conductivity (about 10 \% of total), 
we think that the lattice thermal conductivity 
is nearly equal between the two.
On the contrary, the fully Li-occupied oxide LiCoO$_2$
shows 3-4 times higher $\kappa$ than Na$_{0.77}$MnO$_2$.
This clearly indicates that the A-ion deficiency effectively 
reduces $\kappa$.

Let us fit the data with the phonon-scattering theory \cite{Callaway1,Callaway2}.
The lattice thermal conductivity is written as 
\begin{equation}
 \kappa_{\rm ph} = \frac{k_B}{2\pi^2 v} \left(\frac{k_B}{\hbar}\right)^3  T^3
  \int_0^{\Theta_D/T} \frac{x^4 e^x}{(e^x-1)^2}\ \tau\ dx
\end{equation}
where $\Theta_D$ is the Debye temperature, 
and $v$ is the sound velocity, and $x=\hbar\omega/k_BT$.
The total scattering rate $\tau^{-1}$ is given as the sum of 
three scattering rates as
\begin{equation}
\tau^{-1} = \tau_{pd}^{-1} + \tau_{ph-ph}^{-1} + \tau_{0}^{-1}
=a \omega^4 + b \omega^2 + v/L
\end{equation}
where $\tau_{pd}^{-1}$, $\tau_{ph-ph}^{-1}$ and $\tau_{0}^{-1}$ 
are the scattering rates for the point-defect scattering, 
the phonon-phonon scattering, and the boundary scattering, respectively.
For a phonon frequency $\omega$, the three scattering rates are written as 
$a\omega^4$, $b\omega^2$ and $v/L$, 
where $a$, $b$ and $L$ are characteristic parameters.
According to Klemens \cite{Klemens}, 
the coefficient $a$ is calculated as 
\begin{equation}
a = \Omega_{0} \Sigma f_i (1-M_i/M)^2 /4 v^3,
\end{equation}
where $\Omega_0$ is the unit cell volume,
$M_i$ and $f_i$ are the mass and the fraction of an atom, 
and $M = \Sigma f_i M_i$ is the average mass. 
Then $a$ for Na$_{0.07}$MnO$_2$ can be calculated 
only by assuming that the Na ion and the vacancy make 
a solid solution,
and $a$ is calculated to be zero for LiCoO$_2$.
As discussed in the previous paper \cite{takahata}, 
the temperature dependence of $\kappa$ indicates 
negligibly small phonon-phonon scattering,
and we safely assume $b=0$.
Consequently $v$ and $L$ are to be determined through fitting.
Since $v$ is related to $\Theta_D$ as
\begin{equation}
\Theta_D
  =\frac{\hbar v}{k_B}(6\pi^2N)^{\frac{1}{3}}, 
\end{equation}
and thus we employed $\Theta_D$ instead of $v$ as a fitting parameter
($N$ is the number of atoms per unit cell).

\begin{figure}
 \includegraphics[width=8cm,clip]{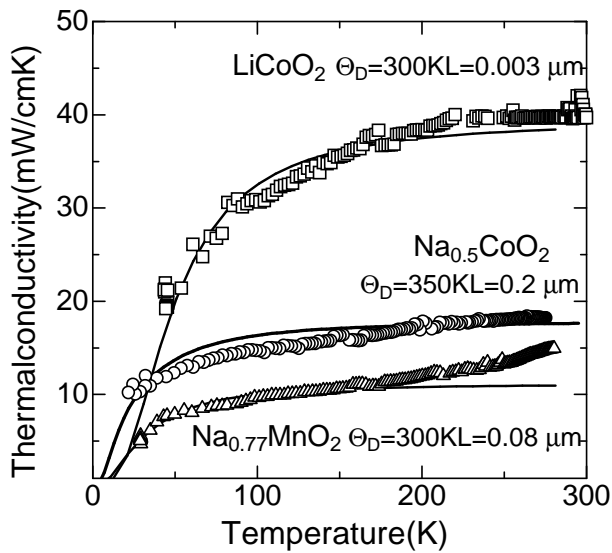}
 \caption{
 The thermal conductivity of A$_x$BO$_2$ type layered oxides plotted as 
 a function of temperature.
 The theoretical fitting by the phonon-scattering theory is shown by the 
 solid curves.
 Note that the theoretical curves are functions as the Debye temperature
 $\Theta_D$ and the boundary scattering length $L$ (see text). 
 }
\end{figure}

The data fitting is satisfactory, as displayed in Fig.2 
by the solid curves.
$\Theta_D$'s are fitted to be 300 K for 
Na$_{0.77}$MnO$_2$ and LiCoO$_2$, 
which are nearly equal to the value of 350 K for NaCo$_2$O$_4$ 
estimated from the specific-heat measurement \cite{ando}.
$L$'s are fitted to be 0.2, 0.08, 0.003 $\mu$m
for NaCo$_2$O$_4$, Na$_{0.77}$MnO$_2$ and LiCoO$_2$, respectively.
It should be noted that $L$ of LiCoO$_2$ is much shorter than the others.
In this compound, we set $a=b=0$ in Eq. (2),
which means that only the boundary scattering dominates $\kappa$.
Then $L$ is reduced to be an ``average'' phonon mean free path $\ell$
in the sense of the classical argument $\kappa = Cv\ell/3$,
where $C$ is the phonon specific heat.
From the very beginning, $L$ is introduced as 
a phenomenological parameter, 
and cannot be evaluated from the first principle.
The physical meaning of $L$ is to be explored theoretically,
as other researchers also pointed out \cite{Chen,Simkin}.

In summary, polycrystalline samples of 
Na$_{0.77}$MnO$_2$ and LiCoO$_2$ are prepared,
and the thermal conductivity is measured from 15 to 280 K.
We have successfully fitted the experimental data 
with the phonon-scattering theory, 
and have found that the reduction of the thermal conductivity is 
mainly determined by the vacancy (point defect), 
not by the layered structure.

The authors appreciate Y. Iguchi, A. Satake, 
W. Kobayashi and R. Kitawaki for collaboration.



\end{document}